\documentclass[twocolumn,showpacs,showkeys,superscriptaddress]{revtex4}
\usepackage{graphicx}
\usepackage{bm}
\usepackage{color}
\usepackage{amsmath}
\usepackage{natbib}
\usepackage{latexsym}

\begin{document}

{\raggedleft {\it Accepted for publication in Physical Review Letters}}
\\

\title{Thermal-inertial effects on magnetic reconnection in relativistic pair plasmas}
\author{Luca Comisso}
\email{luca.comisso@polito.it}
\affiliation{Dipartimento Energia, Politecnico di Torino, Corso Duca degli Abruzzi 24, 10129, Torino, Italy.}
\affiliation{Istituto dei Sistemi Complessi - CNR, Via dei Taurini 19, 00185, Roma, Italy.}
\author{Felipe A. Asenjo}
\email{felipe.asenjo@uai.cl}
\affiliation{Facultad de Ingenier\'{\i}a y Ciencias, Universidad Adolfo Ib\'a\~nez, Santiago, Chile.}
\affiliation{Departamento de Ciencias, Facultad de Artes Liberales,
Universidad Adolfo Ib\'a\~nez, Santiago, Chile.}


\begin{abstract}
The magnetic reconnection process is studied in relativistic pair plasmas when the thermal and inertial properties of the magnetohydrodynamical fluid are included.
We find that in both Sweet-Parker and Petschek relativistic scenarios there is an increase of the reconnection rate owing to the thermal-inertial effects, both satisfying causality. To characterize the new effects we define a thermal-inertial number which is independent of the relativistic Lundquist number, implying that reconnection  can be achieved even for vanishing resistivity as a result of only thermal-inertial effects.
The current model  has fundamental importance for relativistic collisionless reconnection, as it constitutes the simplest way to get reconnection rates faster than those accessible with the sole resistivity.
\end{abstract}

\pacs{52.27.Ep; 52.27.Ny; 52.30.Cv; 52.35.Vd}
\keywords{Magnetic reconnection, Relativistic plasmas, Thermal-inertial effects}

\maketitle

Magnetic reconnection is a fundamental plasma process which is widely believed to play a key role in many phenomena occurring in laboratory, space and astrophysical plasmas. Most of the progress in the theory of magnetic reconnection has been done in the non-relativistic regime \cite{PriFor_2000,Bis_2000}. However, in recent years it has been recognized the importance of reconnection processes in magnetically dominated environments, where special relativistic effects have to be considered \cite{Uzdensky_2011,Hoshino_2012}. Indeed, in these environments the magnetic energy density $B^2/8 \pi$ largely exceeds the rest mass energy density $mnc^2$, and thus the speed of the Alfv\'en wave $v_A=cB/(4\pi mn c^2+B^2)^{1/2}$ approaches the speed of light $c$. In particular, relativistic reconnection is extremely important in pair (electron-positron) plasmas such as those in pulsar magnetospheres \cite{Contopoulos_2007,UzdSpit_2014}, pulsar winds \cite{Coroniti_1990,LyuKirk_2001}, soft gamma-ray repeaters \cite{TD_95,Lyutikov_2003}, jets from gamma-ray bursts \cite{Drenkhahn_2002,McKinney_2012} and from active galactic nuclei \cite{Jaroschek_2004,Giannios_2009}.

In spite of the fact that relativistic magnetic reconnection is becoming increasingly important in many aspects of modern astrophysics, only a few theoretical studies on the fundamental physics have been done. The problem of the relativistic generalization of the classical Sweet-Parker and Petschek reconnection models was approached for the first time by Blackman and Field \cite{Blackman_1994}, who argued that because of Lorentz contraction the inflow velocity of the reconnecting magnetic field is greatly enhanced and may approach to the speed of light. Their conclusion was confirmed by Lyutikov and Uzdensky \cite{LyutUzd_2003} for the relativistic Sweet-Parker scenario. On the contrary, a subsequent analysis by Lyubarsky \cite{lyu} showed that the reconnection inflow remains sub-relativistic in both scenarios. These pioneer works were followed by a study of the relativistic Petschek-type shock with pressure anisotropy \cite{TenBarge_2010}, and by resistive relativistic magnetohydrodynamic (RMHD) simulations which seemed to be more consistent with Lyubarsky's theory \cite{Watanabe2006,Zenitani_2010,Takahashi_2011}.

It is important to point out that all previous theoretical models of relativistic reconnection were developed in the framework of resistive RMHD. However, collisionless effects can significantly affect the reconnection process and their investigation in the relativistic regime is an open problem in astrophysics and fundamental physics. As a contribution towards the clarification of this point, here we extend the previous relativistic reconnection models by considering also thermal and inertial effects in pair plasmas. For this purpose we adopt a relativistic magnetohydrodynamical theory derived from first principles from a two-fluid pair plasma, and we analyze the magnetic reconnection process in the Sweet-Parker and Petschek configurations. We find that in both scenarios the thermal-inertial effects play an essential role which bring new contributions to the reconnection process as compared to the purely resistive case.

{\it Generalized RMHD equations.} A set of equations for a RMHD pair plasma has been recently derived by Koide~\cite{koide}. These equations represent a generalization to the previous simpler models \cite{lich,Anile_1989}, since they are derived in a systematic and rigorous way from the equations of a two-fluid plasma.
For a RMHD pair plasma with density $n$, normalized four-velocity $U^\mu$ (such that $U_\mu U^\mu=\eta_{\mu\nu}U^\mu U^\nu=-1$), normalized four-current density $J^\mu$, and a metric signature $\eta_{\mu\nu}=(-1,1,1,1)$, the generalized RMHD equations  \cite{koide} are composed by the continuity equation
\begin{equation}\label{con1}
  \partial_\mu\left(n U^\mu\right)=0\, ,
\end{equation}
the generalized momentum equation
\begin{eqnarray}\label{mom}
\partial_\nu\left[h U^\nu U^\mu+\frac{h}{4 n^2e^2}J^\nu J^\mu\right]=-\partial^\mu p+J_\nu F^{\mu\nu}\, ,
\end{eqnarray}
and the generalized Ohm's law
\begin{eqnarray}\label{ohm}
&&\frac{1}{4 n e}\partial_\nu\left[\frac{h}{n e}\left(U^\mu J^\nu+J^\mu U^\nu\right)\right]\nonumber\\
&&\qquad\qquad=U_\nu F^{\mu\nu}-\eta\, c\left[J^\mu+U_\alpha J^\alpha U^\mu(1+\Theta)\right] .
\end{eqnarray}
Here, $h$ is the enthalpy of the RMHD pair plasma and $e$ stands for the electron charge. The pressure is represented by $p$, whereas $F^{\mu\nu}$ is the electromagnetic tensor field. The resistivity can be recognized as $\eta$, and $\Theta$ is the thermal energy exchange rate from negative to positive charged fluids.
In the above model the variation of enthalpy and pressure between the positively and negatively charged fluids are considered as negligible.

For a pair plasma, Koide \cite{koide} obtained that  $\Theta=2\varpi[(U_\mu J^\mu)^2+J_\mu J^\mu]/[4n^2e^2-(U_\mu J^\mu)^2]$, where $\varpi$ is the coefficient of thermalized energy due to the friction of the fluids.
In general we can define a thermal function $f = f(T)=h/(mn c^2)$ depending only on the temperature $T$.
For the simplest calculation of the enthalpy $h$ of a relativistic plasma in thermal equilibrium \cite{mahajanU}, $f=K_3(m c^2/k_B T)/K_2(m c^2/k_B T)$, where $K_n$ is the modified Bessel function of order $n$ and $k_B$ is the Boltzmann constant. Collision effects have not been taken into account to obtain $h$, and for the purposes of the current work we will consider it as the first approximation to a more general form of the enthalpy.
For relativistically hot plasmas $k_B T\gg mc^2$, so that $f\approx 4 k_BT/mc^2$ and $h\approx 4p$, with the plasma pressure $p=nk_B T$.

The previous set of equations must be complemented by Maxwell's equations
\begin{equation}\label{Maxw}
  \partial_\nu F^{\mu\nu}= 4\pi J^\mu\, ,\qquad \partial_\nu F^{*\mu\nu}=0\, ,
\end{equation}
where $F^{*\mu\nu}$ is the dual tensor density of the electromagnetic tensor.

In this generalized RMHD model, the inertial effects, proportional to $h$, modify the momentum equation and Ohm's law. In Eq.~\eqref{mom}, the current inertia effects arise from the left-hand side. On the other hand, in the left-hand side of Eq.~\eqref{ohm} the thermal electromotive effects appear as inertial effects corrections.

\begin{figure}[h!]
\begin{center}
\includegraphics[bb = 0 0 314 246, width=8.6cm]{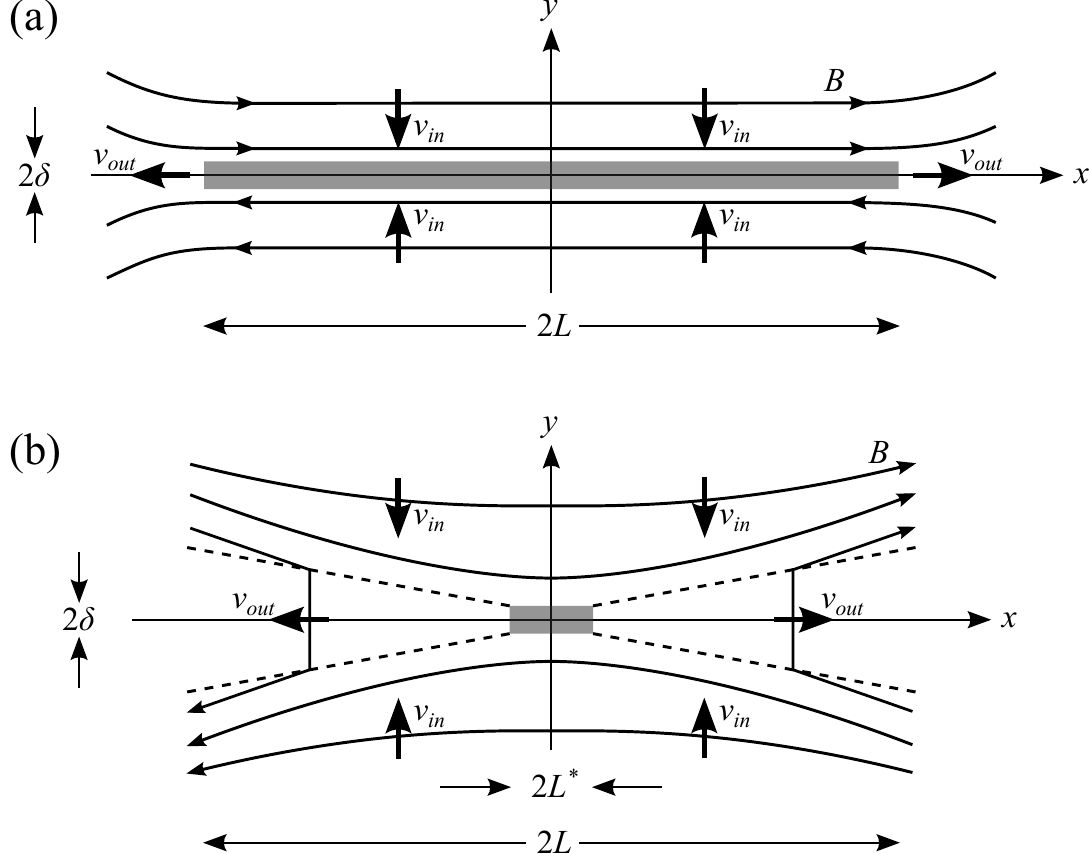}
\end{center}
\caption{Geometry of the (a) Sweet-Parker and (b) Petschek configurations. The magnetic diffusion regions are shaded, while slow mode shocks are indicated by dashed lines.}
\label{fig1}
\end{figure}

{\it Sweet-Parker configuration.} As in the classical Sweet-Parker theory, in our analysis we consider an elongated magnetic diffusion region (with length $2 L$ and width $2 \delta \ll 2 L$) which lies between opposite directed magnetic field lines, as shown in Fig. \ref{fig1}(a). Outside the diffusion region the plasma is highly ideal, such that the frozen-in flux condition holds. The magnetic field and the plasma velocity are in the $xy$-plane, with the origin $(0,0)$ representing a stagnation point for the flow. We consider a steady state and we assume that all the physical quantities are independent of $z$. Furthermore, the magnetic field upstream of the diffusion region, indicated with $B_0$, is in the $x$-direction and of equal strength on opposite sides of the layer.

Since we are considering magnetically dominated environments, the upstream plasma pressure can be neglected compared to the magnetic pressure, and thus the pressure balance across the layer gives $p=B_0^2/ 8\pi$, where $p$ is the plasma pressure in the center of the diffusion region.
Besides, close to the neutral line we find that $E^x\approx 0 \approx E^y$ and $B^z \approx 0$, while $J^0=0\approx J^x$, and $v^y\approx0 \approx v^z$, implying that $U_\mu J^\mu \approx 0$. Therefore there is no contribution from the thermal energy exchange rate between the charged fluids.
Then, from the momentum equation \eqref{mom} along the neutral line we have
\begin{equation}\label{mom2}
  \partial_x\left(h \gamma_{out} \frac{v_{out}}{c} U^\mu\right)=-\partial^\mu p+J_\nu F^{\mu\nu}\, ,
\end{equation}
where $v_{out}$ is the outflow velocity with its respective Lorentz factor $\gamma_{out}=(1-v_{out}^2/c^2)^{-1/2}$. The current inertia do not play any role in Eq.~\eqref{mom2} due to the direction of the current density in the reconnection layer.
In the $x$-direction, Eq.~\eqref{mom2} implies that
\begin{equation}\label{ph}
h \gamma_{out}^2 \frac{v^2_{out}}{c^2} + p = -L  {J^z B^y}  \sim  \frac{ B_0^2}{4\pi}  =2p\, ,
\end{equation}
where we have used Maxwell's equation \eqref{Maxw} to estimate the current density in the $z$-direction $J^z \approx -B_0/(4\pi\delta)$, and the flux conservation for the outflow magnetic field strength $B^y\sim B_0 \delta /L$. Eq.~\eqref{ph} suggests that $h \gamma_{out}^2 {v^2_{out}}/{c^2}\sim p$. Since for relativistically hot plasmas $h \approx 4 p$, it follows that the outflow velocity is mildly-relativistic with  ${v_{out}}\sim c$ and $\gamma_{out}\sim 1$, as shown in the purely resistive case by Lyubarsky \cite{lyu}.

In the diffusion region, the generalized Ohm's law \eqref{ohm} becomes
\begin{equation}\label{Ohm2}
\eta c J^\mu + \frac{1}{4 n e}\partial_\nu\left[\frac{h}{n e}\left(U^\mu J^\nu+J^\mu U^\nu\right)\right] =U_\nu F^{\mu\nu} \, ,
\end{equation}
since there is no contribution from the thermal energy exchange rate close to the neutral line. Following the Sweet-Parker scheme, we find that the thermal electromotive effects in the reconnection layer are estimated as $\partial_x\left[{h}\left(U^\mu J^x+J^\mu U^x\right)/({n e})\right] \approx  \partial_x\left[{h}J^\mu U^x /{n e}\right]\sim h\gamma_{out} v_{out} J^\mu/(L n e c)$.
 Then, Eq.~\eqref{Ohm2} can be reduced to
\begin{equation}\label{OhmS}
  (\eta+\beta)J^\mu=\frac{1}{c}U_\nu F^{\mu\nu}\, ,
\end{equation}
where we have introduced a thermal-inertial parameter defined as
\begin{equation}\label{}
  \frac{h\gamma_{out} v_{out}}{4 n^2e^2 L c^2}\sim \frac{h}{4 n^2 e^2 L c}=\frac{\pi f \lambda_e^2}{L c}=\beta \, ,
\end{equation}
with $\lambda_e=c/{\omega_p}$ indicating the electron skin depth and $\omega_p$ the electron plasma frequency. We can see that thermal electromotive effects introduce an inertial correction to Ohm's law, whose $y$- and $z$-components yield $J^y = 0$ and
\begin{equation}\label{ohmsweet}
  (\eta+\beta)J^z= \frac{1}{c}E^z\, .
\end{equation}

In a steady state two-dimensional configuration the out-of-plane electric field is uniform by virtue of Maxwell's equation \eqref{Maxw}. Hence, $E^z$ in Eq.~\eqref{ohmsweet} can be evaluated from Ohm's law in the ideal region just upstream of the reconnection layer, which gives $E^z=v_{in} B_0/c$. Moreover, balancing the inflowing electromagnetic energy with the energy outflow, i.e. $L E^z B_0 c / 4\pi \sim \delta h v_{out}$, we have that
$ \delta  \sim  (v_{in}/v_{out}) L \sim  (v_{in}/c) L$,
implying that the plasma is approximately incompressible. From this relation, using that $\delta \approx B_0/(4 \pi J^z)$ and eliminating $J^z$ through Eq.~\eqref{ohmsweet}, we find the reconnection rate
\begin{equation}\label{}
  \frac{v_{in}}{c}  \sim  \sqrt{\frac{1}{S}+\frac{\beta c}{4\pi L}}\, ,
\end{equation}
where $S= 4\pi L/\eta c \gg 1$ is the relativistic Lundquist number.
Thermal-inertial effects contribute to the reconnection layer width and to the reconnection rate through the ``thermal-inertial number''
\begin{equation}\label{thermalinertialnumber}
  \frac{4\pi L}{\beta c} = \frac{4}{f d_e^2}\, ,
\end{equation}
where $d_e={\lambda_e}/{L}$ is the dimensionless electron inertial length. We can see that $\beta c$ introduces relativistic effects trough the enthalpy $h$ or the thermal function $f$, which depends on the ratio between the particles rest mass energy and the relativistic temperature. For non-relativistic plasmas, $f=1$. In a general case, $f\geq 1$ always (for relativistically hot plasmas $f \gg 1$), but since the diffusion region is supposed to be localized, i.e. $d_e \ll 1$, we expect $4/f d_e^2 > 1$. Hence, although the thermal-inertial effects contribute to increase significantly the reconnection rate with respect to the purely resistive case \cite{lyu}, the inflow velocity of the reconnecting magnetic field is expected to remain sub-relativistic.

Thermal-inertial effects were not considered in previous analytic treatments \cite{Blackman_1994,LyutUzd_2003,lyu} whose purpose was to formulate a relativistic generalization of the Sweet-Parker reconnection model. However, we would like to stress that thermal-inertial effects by themselves allow magnetic reconnection to take place. These effects become relevant if $\beta \gtrsim \eta$, namely when the thermal-inertial layer width $\delta_{ti} \sim \sqrt{f } \, \lambda_e /2$ is of the same order or larger than the resistive layer width $\delta_{\eta} \sim S^{-1/2} L$. In particular, for hot plasmas in which $f  \approx  4k_B T/ m c^2$, this condition can be written as $k_B T / mc^2 \gtrsim {1}/ S d_e^2$, which gives a relation between the thermal to electron rest mass energy ratio and the non-idealness of the plasma.

{\it Petschek configuration.} It will be shown that thermal-inertial effects play a key role also in the Petschek scenario in which a relatively short diffusion region (of length $2L^* \ll 2L$) act as a source for two pairs of slow mode shocks, as shown in Fig. \ref{fig1}(b). The shocks stand in the flow when a steady state is reached, marking the boundaries of the outflow regions. In this scenario, the magnetic energy conversion takes place not only in the diffusion region, but also across the slow mode shocks.

In order to evaluate the reconnection rate in this configuration, we need to formulate the jump relations at the shocks for the relativistic pair plasma fluid. For this purpose we observe that the momentum equation \eqref{mom} can be written in the form of the conservation law
\begin{equation}
  \partial_\nu T^{\mu\nu}=0\, ,
\end{equation}
where the total energy-momentum is $T^{\mu\nu}=T^{\mu\nu}_f+T^{\mu\nu}_{em}$, with the  energy-momentum tensor of the fluid
\begin{equation}
  T_f^{\mu\nu} = h U^\mu U^\nu + \frac{h}{4n^2e^2}J^\mu J^\nu + p\, \eta^{\mu\nu}\, ,
\end{equation}
and the electromagnetic energy-momentum tensor
\begin{equation}
  T_{em}^{\mu\nu}=\frac{1}{4\pi}F^{\mu\beta}{F^\nu}_\beta-\frac{1}{16\pi}F^{\alpha\beta}F_{\alpha\beta}\eta^{\mu\nu}\, .
\end{equation}
Then, in a reference frame in which the shock front is at rest, from the conservation of energy and momentum fluxes we get
\begin{equation}\label{Pes1}
  \rho_1\gamma_1^2 \frac{v_1}{c}+\frac{1}{4\pi} B_{t1}E_t=h_2\gamma_2^2\frac{v_{n2}}{c}+\frac{1}{4\pi} B_{t2}E_t\, ,
\end{equation}
\begin{eqnarray}\label{Pes2}
  \rho_1\gamma_1^2 \frac{v_1^2}{c^2}+\frac{\rho_1}{4n^2e^2}J_1^2+\frac{1}{8\pi} B_{t1}^2&=&h_2\gamma_2^2\frac{v_{n2}^2}{c^2}+\frac{h_2}{4n^2e^2}J_{n2}^2\nonumber\\
&&+p_2+\frac{1}{8\pi} B_{t2}^2\, ,
\end{eqnarray}
\begin{equation}\label{Pes3}
-\frac{1}{4\pi}B_n B_{t1}=h_2\gamma_2^2 \frac{v_{n2} v_{t2}}{c^2}+\frac{h_2}{4n^2e^2}J_{n2} J_{t2}-\frac{1}{4\pi}B_n B_{t2}\, ,
\end{equation}
where the subscripts $1$ and $2$ refer to the upstream and downstream flows respectively, and the subscripts $n$ and $t$  refer to the normal and tangential components of the fields with respect to the shock plane.
Also we assume that the flow is cold upstream ($f_1=1, \, h_1=m n c^2=\rho_1$) and hot downstream ($h_2=4 p_2$).

The set of Eqs.~\eqref{Pes1}-\eqref{Pes3} contains the corrections of the current inertia effects. In the same fashion, from Ohm's law \eqref{ohm}, we have in the upstream flow (with resistivity $\eta=0$)
\begin{equation}\label{ohmu}
  \frac{1}{4n e^2 c \gamma_1}\partial_n\left(\frac{\rho_1}{n}\gamma_1 v_1 J_1\right)=E_t-\frac{v_1}{c} B_{t1}\, ,
\end{equation}
where $\partial_n$ is the space-derivative along the normal direction of the shock plane. For the downstream flow, Ohm's law becomes
\begin{eqnarray}\label{ohmd}
 &&\frac{1}{4n e^2 c \gamma_2}\partial_\parallel\left(\frac{h_2}{n}\gamma_1 v_1J_{t2}+\frac{h_2}{n}J_1\gamma_2 v_{t2}\right)\nonumber\\
&&+ \frac{1}{4 n e^2 c \gamma_2}\partial_n\left(\frac{h_2}{n}J_1\gamma_2 v_{n2}+\frac{h_2}{n}J_{n2}\gamma_2 v_{t2}\right)\nonumber\\
&&=E_t+\frac{v_{t2}}{c} B_n-\frac{v_{n2}}{c}B_{t2}\, ,
\end{eqnarray}
where $\partial_\parallel$ is the space-derivative in the tangential direction to the shock plane.

This model have to be solved in the Petschek scenario in which the standing slow mode shocks are of the switch-off type, namely with  $B_{t2}=0$. The currents can be estimated from Maxwell's equations \eqref{Maxw}, so that $4\pi J_{n2}\sim \partial_\parallel B_{t2}= 0$, $4\pi J_{t2}\sim -\partial_\parallel B_{n}$, and $4\pi J_{1}\sim -\partial_\parallel B_{t1}$.
The inertial effects introduce nonlinear terms that make a solution difficult to be foreseeing. However, we must notice that the magnetic field gradients along the shock plane are in general negligible, so that a uniform shock plane can be formed at large distances. This is indeed confirmed by simulations of electron-positron reconnection \cite{zenitani_2009}. Thereby, we can assume $\partial_\parallel B_{n}\approx 0\approx \partial_\parallel B_{t1}$, giving that $J_{1}\approx0\approx J_{t2}$ on the shock plane. This implies that the previous Eqs. \eqref{Pes1} - \eqref{Pes3} reduce to
\begin{equation}\label{Pes1b}
  \rho_1\gamma_1^2 v_1+\frac{c}{4\pi} B_{t1}E_t=h_2\gamma_2^2v_{n2}\, ,
\end{equation}
\begin{eqnarray}\label{Pes2b}
  \rho_1\gamma_1^2 v_1^2+\frac{c^2}{8\pi} B_{t1}^2=h_2\gamma_2^2v_{n2}^2+c^2p_2\, ,
\end{eqnarray}
\begin{equation}\label{Pes3b}
-\frac{c^2}{4\pi}B_n B_{t1}=h_2\gamma_2^2 v_{n2} v_{t2}\, ,
\end{equation}
whereas Eqs.~\eqref{ohmu} and \eqref{ohmd} combine to give
\begin{equation}\label{ohme}
 c E_t=v_1 B_{t1}=-v_{t2} B_n\, .
\end{equation}

The above system of equations is the same one found by Lyubarsky \cite{lyu}. The thermal-inertial effects play a negligible role across the switch-off shocks, however we will show that they are crucial in the diffusion region.
Eqs.~\eqref{Pes1b}, \eqref{Pes3b} and \eqref{ohme} can be combined to find that
\begin{equation}\label{}
 \frac{v_1^2}{c^2}=\frac{B_n^2}{4\pi\gamma_1^2\rho_1+B_{t1}^2}=\frac{\tan^2\theta}{1+{1}/(\sigma_1 \cos^2\theta)}\, ,
\end{equation}
showing that the velocity of the upstream flow is the Alfv\'en velocity. Here we have indicated with $\theta$ the angle between the magnetic field and the shock plane, so that $B_{t1}=B_1 \cos\theta$ and $B_n=B_1\sin\theta$, while $\sigma_1=B_1^2/(4\pi\gamma_1^2 \rho_1)$  is the magnetization parameter upstream to the shock.
In magnetically dominated environments $\sigma_1 \gg 1$, therefore, assuming $\theta < \pi/4$, the upstream velocity becomes $v_1\approx c\tan\theta$. The other variables can also be solved in terms of $\theta$ obtaining \cite{lyu}  $v_{t2}\approx -c+c\sec^2\theta  / 2\sigma_1$, $v_{n2}\approx c\tan\theta \sec^2\theta  / 2\sigma_1$, $\gamma_2\approx \sqrt{\sigma_1}\cos\theta$ and $p_2\approx B_1^2 \cos^2\theta/8\pi$. Thus, the outflow velocity is ultra-relativistic ($\gamma_{out} \gg 1$) and forms an angle $\varphi$ with the slow mode shock that is inversely proportional to the magnetization parameter $\sigma_1$, since $\tan\varphi \approx - \tan\theta \sec^2\theta  / \sigma_1$ for $\sigma_1 \gg 1$.

Petschek's regime is almost-uniform as it assumes that the magnetic field in the inflow region is a small perturbation to a uniform magnetic field $B_0$. Furthermore, it is assumed that the magnetic field changes mainly within the diffusion region, whereas outside it is irrotational and $\bf{B}=\nabla\psi$, so that $\partial_\mu \partial^\mu \psi = 0$ in a steady state. Following a standard procedure \cite{PriFor_2000}, the magnetic field in the upper inflow region can be evaluated adding $B_0$ to the magnetic field obtained by solving Laplace's equation in the upper half-plane with appropriate boundary conditions. To lowest order, neglecting the inclination of the shocks, these conditions are $B^y(x,0)=-2B_n$ for $-L \le x \le -L^*$, $B^y(x,0)=2B_n$ for $L^* \le x \le L$, and a magnetic field perturbation that vanish at infinity and at the diffusion region. Then, the magnetic field just upstream of the diffusion region is
\begin{equation}\label{fieldorig}
B^x(0,\delta) = B_0 \left( {1 - \frac{4v_{in}}{\pi c}\ln \frac{L}{L^*}} \right)\, .
\end{equation}
The length $L^*$ can be estimated from the Sweet-Parker relations for the diffusion region and flux conservation $v^y(0,\delta) B^x(0,\delta) = v_{in} B_0$. Therefore we get
\begin{equation}
L^* \sim  \frac{(\eta c + \beta c)}{4 \pi} \left( {\frac{c}{v_{in}}} \right)^2  \, .
\end{equation}
As in the classical Petschek model, the mechanism hangs itself when the magnetic field immediately upstream of the diffusion region becomes too small. The maximum reconnection rate occurs for $B^x(0,\delta)/B_0 \sim 1/2$, so that from Eq. \eqref{fieldorig} we obtain
\begin{equation}\label{vinPes}
 \left. {\frac{v_{in}}{c}} \right|_{\max } \sim \frac{\pi}{8}\left[\ln\left(\frac{4\pi L}{\eta c + \beta c}\right)\right]^{-1}\, .
\end{equation}
This relation shows that the reconnection rate can be high also for vanishing resistivity because of thermal-inertial effects. These effects become relevant under the same conditions of the relativistic Sweet-Parker scenario, so that in hot tenuous plasmas they can be responsible for a substantial increase of the reconnection rate.

In both models studied here, the enhancement of the reconnection rate with respect to the purely resistive case can be understood by recognizing that $\beta$ plays the role of a ``thermal-inertial resistivity'' that limits the response of the electrons/positrons to the reconnection electric field. Thereby, the effective resistivity can be significantly heightened, leading to a diffusion region with a smaller aspect ratio that can sustain fast magnetic field line merging. The thermal-inertial resistivity behaves as $\beta\propto n^{-1}$, which is consistent with the results of recent numerical simulations of pair plasma reconnection \cite{zenitani_2009,Bessho_2012}, where it was found that the reconnection rate becomes higher as the particle number density decreases.

Thermal-inertial effects allow the decoupling of the plasma motion from that of the magnetic field lines also in the non-relativistic limit, but in relativistically hot plasmas they are enhanced due to the increase of the thermal function $f$. Furthermore, we observe that since the information propagation velocity is given by the ``head velocity'' $v_h = {\lim_{\omega \to \infty}} \omega/k$ \cite{koide_2011}, which is always $\leq c$ in generalized RMHD, the thermal-inertial effects, as well as the classical resistive effects, always satisfy causality.

{\it Conclusions.} Using an improved set of equations for RMHD plasmas, in which collisionless effects are considered, we have found robust features of the thermal and inertial effects on the magnetic reconnection process in relativistic pair plasmas. In both Sweet-Parker and Petschek configurations the thermal-inertial effects introduce new corrections to the reconnection rates. They provide an effective mechanism for the reconnection of magnetic field lines in the relativistic regime, which also works for vanishing resistivity. We have defined a thermal-inertial number \eqref{thermalinertialnumber} that characterizes the strength of these effects. This new number depends on a thermal function $f$, varying according to the temperature of the plasma, and on the electron inertial length $\lambda_e$, which is inversely proportional to the square root of the electron number density. Thereby, the thermal-inertial effects become relevant in hot tenuous plasmas. We have shown that if the thermal-inertial layer width $\delta_{ti} \sim \sqrt{f} \, \lambda_e /2$ exceeds the resistive layer width $\delta_{\eta} \sim S^{-1/2} L$, the reconnection process enters into the collisionless regime in which thermal-inertial effects dominate. As a result, the reconnection rate in the relativistic regime can be much higher than previously predicted by purely resistive RMHD models.

{\it Acknowledgments.} The authors are grateful to Seiji Zenitani and an anonymous referee for valuable comments. FAA thanks to CONICyT-Chile for Funding N$^o$ 79130002.

\end{document}